\documentclass[conference]{IEEEtran}
\IEEEoverridecommandlockouts
\usepackage{amsthm}
\usepackage{makecell}
\usepackage{ifpdf}
\usepackage{cite}
\usepackage{placeins}
\usepackage[pdftex]{graphicx}
\usepackage{array}
\usepackage{caption}
\usepackage{subcaption}
\usepackage{fixltx2e}
\usepackage{stfloats}
\usepackage{amsthm}
\usepackage{amsmath}
\usepackage{bbm}
\usepackage{amsfonts}
\usepackage{amssymb}
\usepackage[T1]{fontenc} 
\usepackage[cmintegrals]{newtxmath}
\usepackage{bm} 
\usepackage[all]{xy}
\usepackage{enumitem}
\usepackage{float}
\usepackage[usenames, dvipsnames]{color}
\usepackage{ifpdf}
\usepackage{cite}
\usepackage{array}
\usepackage{mathtools}

\makeatletter
\def\widebreve{\mathpalette\wide@breve}
\def\wide@breve#1#2{\sbox\z@{$#1#2$}%
	\mathop{\vbox{\m@th\ialign{##\crcr
				\kern0.08em\brevefill#1{0.8\wd\z@}\crcr\noalign{\nointerlineskip}%
				$\hss#1#2\hss$\crcr}}}\limits}
\def\brevefill#1#2{$\m@th\sbox\tw@{$#1($}%
	\hss\resizebox{#2}{\wd\tw@}{\rotatebox[origin=c]{90}{\upshape(}}\hss$}
\makeatletter

\usepackage{url}
\usepackage{mathtools}
\usepackage[dvipsnames]{xcolor}
\usepackage{multirow}
\usepackage{mwe} 
\usepackage{times}
\usepackage{wasysym}
\usepackage{xr}
\usepackage{epsfig}
\usepackage{multirow}
\usepackage{tabularx}
\usepackage{graphicx}
\usepackage{color}
\usepackage{xspace}
\usepackage{thumbpdf}
\usepackage{listings}
\usepackage{verbatim}
\usepackage[font=small]{caption}
\usepackage{outlines}
\usepackage{textgreek}
\usepackage{booktabs}
\usepackage{amsfonts}
\usepackage{colortbl}
\usepackage{multicol}
\usepackage{multirow}
\usepackage{balance}
\usepackage{tabularx}
\usepackage{makecell}
\usepackage{array}
\usepackage[normalem]{ulem}
\usepackage{mathtools,amssymb}
\usepackage{float} 
\usepackage{booktabs} 
\usepackage{wrapfig}
\usepackage{subfloat}

\usepackage{diagbox}
\usepackage{cleveref}
\usepackage[letterpaper, left=0.625in, right=0.625in, top=0.75in, bottom=1in]{geometry}
\usepackage[]{algorithm2e}

\theoremstyle{definition}

\usepackage{amsthm}

\newcommand{\no}{\nonumber}

\usepackage[noend]{algpseudocode}

\usepackage{diagbox}

\theoremstyle{definition}

\begin{document}
\title{Optimal Routing and Link Configuration for Covert Heterogeneous Wireless Networks in the Presence of a Friendly Jammer \\}

 \author{%
   \IEEEauthorblockN{Amna Gillani\IEEEauthorrefmark{1},
                     Beatriz Lorenzo\IEEEauthorrefmark{1},
                     Majid Ghaderi\IEEEauthorrefmark{2},
                     Fikadu Dagefu\IEEEauthorrefmark{3}, Justin Kong\IEEEauthorrefmark{3}, and
                     Dennis~Goeckel\IEEEauthorrefmark{1}} \\
   \IEEEauthorblockA{\IEEEauthorrefmark{1}%
                     Department of Electrical and Computer Engineering,
                     University of Massachusetts Amherst,
                     Amherst, MA, 01003 USA\\
                     \{agillani, blorenzo, dgoeckel\}@umass.edu}
    \IEEEauthorblockA{\IEEEauthorrefmark{2}%
                     Department of Computer Science, University of Calgary, Calgary, AB T2N 1N4, Canada \\
                     mghaderi@ucalgary.ca}
   \IEEEauthorblockA{\IEEEauthorrefmark{3}%
                     U.S. Army Combat Capabilities Development Command (DEVCOM), Army Research Laboratory, Adelphi, MD 20783 USA \\
                     \{fikadu.t.dagefu.civ, justin.h.kong2.civ\}@army.mil \\
                     \vspace*{-0.5in}}
    \thanks{This work was supported in part by the National Science Foundation under
ECCS-2148159. Research was also sponsored in part by the Army Research Laboratory and was accomplished under Cooperative Agreement Number W911NF-23-2-0014. The views and conclusions contained in this document are those of the authors and should not be interpreted as representing the official policies, either expressed or implied, of the Army Research Laboratory
or the U.S. Government. The U.S. Government is authorized to reproduce and distribute reprints for Government purposes, not withstanding any copyright
notation herein.

A preliminary version of this work has been accepted for presentation at IEEE VTC-Fall 2026. The present manuscript contains substantial extensions, including additional theoretical development, routing algorithms, performance analysis, and numerical results. This manuscript is currently under review for publication.
}
 }
\maketitle

\begin{abstract}
In modern radio networks, nodes frequently access multiple communication interfaces such as WiFi, cellular, LoRa, and Zigbee. Optimal utilization of such heterogeneous networks (HetNets) at link and network levels is essential for ensuring efficient and secure communication. Some applications require a high level of security, requiring the signal to be completely undetectable. Previous works have considered such covertness, but it often results in limited achievable rates. Physical layer analysis shows that friendly jamming can significantly improve covert data rates, motivating its incorporation into HetNets.  Here, we analyze a scenario where a jammer assists communication in a HetNet in the presence of an adversary attempting signal detection.  We first optimize the physical layer (PHY) for a single link and then incorporate those results into an optimal routing and link configuration approach that accounts for an adversary observing the aggregate signals from all links.  Numerical results demonstrate significant performance gains when compared to alternative approaches.  In fact, the rate observed for the proposed approach is high enough to question the optimality of the low rate design approach employed; we address this concern through revised algorithms and characterize their performance.
\end{abstract}
\vspace*{-0.1in}

\section{Introduction}
\label{sec:introduction}
Heterogeneous networks (HetNets) integrate various radio access technologies, such as WiFi, 5G, and LoRa, to enhance coverage, capacity, and efficiency. Resource allocation across multiple interfaces (i.e., modes) \cite{xu2021survey} and mode selection for HetNets have been extensively studied \cite{feng2020smart, algedir2020energy}. Here, we consider covert communication in HetNets, which requires keeping the presence of the signal hidden from an attentive and powerful adversary \cite{bash2013limits}.

Early work demonstrated the difficulty of the covert communication problem: only $O(\sqrt{n})$ bits can be sent in $n$ channel uses when there is additive white Gaussian noise (AWGN) between the transmitter and each of the intended receiver and the adversary \cite{bash2013limits}; this is often referred to as the square-root law (SRL). Sheikholeslami et al.\cite{covert_routing_2018} proposed a minimum delay and maximum throughput routing scheme for homogeneous covert networks in the presence of multiple adversaries.  We considered the extension of such to HetNets in \cite{gillani2024optimal}.  We proposed a scheme to determine the jointly optimal route and link configuration in polynomial time under an end-to-end covertness constraint.  Our numerical results demonstrate that using multiple modalities significantly improves link and network-level performance. A related study is presented by Kong et al. in \cite{kong2026simultaneous}, which investigates simultaneous use of multiple modalities for covert communication but is limited to single-link scenarios.

However, the data rate for the scenario considered in \cite{gillani2024optimal} is still limited by the strict covertness constraint, especially for longer block lengths due to the SRL \cite{bash2013limits}. 
Lee and Baxley \cite{lee2014achieving,lee2015achieving} showed that a positive covert rate can be achieved on a link when the adversary has uncertainty about the value of their noise power.  A promising method for
introducing such uncertainty in homogeneous networks is demonstrated in \cite{sobers2017covert}, where it is shown that exploiting a friendly jammer can allow for the transmission of $O(n)$ covert bits in $n$ channel uses.  To achieve this, the jammer transmits ``noise'' with a randomly chosen power, hence obfuscating the channel behavior when the transmitter is not present and making it difficult for the adversary to tell whether the measured power variation is due to such randomness or due to the presence of the transmitter \cite{sobers2017covert}. Therefore, here we introduce a jammer in a heterogeneous network and perform joint physical layer (PHY) and routing optimization. In addition to the introduction of a jammer, a key feature of our work is the incorporation of an end-to-end covertness constraint, which captures the true covertness constraint against a powerful adversary that has visibility along the entire route.  This is in contrast to prior work, which, for analytical ease, often enforces a fixed per-link covertness constraint.

The main contributions of this work are as follows:

\begin{itemize}
    \item We optimize single-link covert communications in a HetNet in the presence of a jammer with randomized transmit power. 
    \begin{itemize} 
        \item In the low data-rate regime inherent to covert communication, we linearize the capacity expression and show that, unlike \cite{gillani2024optimal}, maximizing the covert capacity under jamming leads to the selection of a single optimal mode.
        \item For moderate data-rate and the true non-linear capacity expression, we propose a water-filling based solution to optimally allocate the power across modes while adhering to the covertness constraint. 
    \end{itemize}
    \item Leveraging these single-link insights, we propose a method to perform optimal joint routing and link configuration that maximizes capacity while meeting the covertness constraint. We refer to this framework as \emph{Covert Routing Optimization with Jammer (CROP-J)}.
    \begin{itemize}
        \item \textbf{L-CROP-J (Linearized CROP-J):} We propose a polynomial-time covert optimal routing algorithm with a jammer that jointly optimizes the routing and link configuration for multi-hop communications in the presence of a jammer by optimally allocating power across the route while satisfying the end-to-end covertness constraint when the linearized capacity formula is employed.
        \item Motivated by the observation that jamming can enhance the covert capacity and enable operation in a moderate data-rate regime, we extend our analysis to an exact (non-linearized) covert capacity formulation and propose the following algorithms:
        \begin{itemize}
            \item \textbf{S-CROP-J (Single-mode CROP-J):} a mode selection, covert optimal routing algorithm that selects a single optimal mode per link and utilizes the exact capacity expression.
            \item \textbf{WF-Optimal (Water-filling based optimal benchmark):} a benchmark algorithm which performs an exhaustive search over all feasible routes and utilizes water-filling to optimally allocate the power across links while satisfying the end-to-end covertness constraint.
            \item \textbf{WF-CROP-J (Water-filling based CROP-J):} a low-complexity enhancement that applies water-filling over routes obtained from CROP-J, achieving performance between S-CROP-J and WF-Optimal.
        \end{itemize}
    \end{itemize}
\end{itemize}

After reviewing related work in Section \ref{sec: related_work}, Section \ref{sec: model} presents the system model. Section \ref{sec:adversary_analysis} analyzes the adversary model, and Section \ref{sec: Proposed Algorithms} presents the optimal routing and link configuration in a multi-hop network for low and moderate data rates. The numerical results are provided in Section \ref{sec: Numerical results}, and Section \ref{sec:conclusion} concludes the paper.

\section{Related Work}

\label{sec: related_work}  
Most research in covert communication has focused on single-link, single-mode optimization, neglecting the potential of leveraging multiple communication modalities and network-level optimization, even when incorporating more advanced elements such as UAV platforms, multi-antenna jammers, or cooperative interference. For example, UAV-jammer-based covert communication has been examined in \cite{9121687, 9849051, 9536435}, yet these works remain restricted to optimizing a single transmitter-receiver pair. 

To address challenges in heterogeneous networks, a reinforcement learning-based method for optimal route selection was introduced in \cite{kong2024decentralized, kong2025joint} by Kong et al. and in \cite{roknilamouki2025safe} by A. Roknilamouki et al., followed by an extension of the resilient method to node failure by Kim et al. in \cite{10773635}. These works focus on leveraging reinforcement learning to adaptively find optimal routes in more complex and varied network environments, but these do not use multiple modalities simultaneously or consider end-to-end covertness constraint, as done in \cite{gillani2024optimal}. Moreover, in this work, we extend the covert routing in HetNets by incorporating a friendly jammer to enhance the covert capacity. 

The roles of different jammer types, such as informed, uninformed, cooperative, or cognitive jammers, have been explored in \cite{9834682,9120366}, as well as covert communication with a friendly or probabilistic jammer \cite{10236006,10534096}. Covert communication in cognitive radio networks with Poisson distributed jammers or large-scale jamming environment has been studied in \cite{10533214,10143419}, while distributed cooperative jamming architectures have been considered over slow-fading channels in \cite{9442311}. Multi-receiver but still single-transmitter scenarios have been addressed in \cite{9361424}, and multi-UAV systems with informed jammers have been explored in \cite{11214184}.

Moreover, \cite{9360663} investigates mode selection and cooperative jamming for covert communication in D2D-underlaid UAV networks. However, that work focuses solely on selecting the optimal transmission mode for a single D2D link under UAV supervision, without addressing network-level routing or joint optimization across multiple links. The work in \cite{9360663} is abstracted from the PHY and does not present the detailed model-based analysis considered here.  The work in \cite{10256033} and  \cite{9674386} considers jammer selection from multiple jammers for fixed-power jamming in a single mode on a single link.  In contrast, our single-link formulation considers a single jammer with multiple modes, and the additional challenges that this formulation entails, particularly in characterizing the adversary's performance, will be clear from Section \ref{sec:adversary_analysis}.  Finally, both \cite{9360663} and \cite{10256033} consider only single-link scenarios.  Xia et al. \cite{xia2024uav} studied multi-mode selection, data scheduling, and power optimization in a two-hop scenario.

In contrast, a multi-hop network as considered here enhances covertness by distributing transmissions across multiple intermediate nodes, making it more difficult for adversaries to detect or trace the communication path. This approach is particularly valuable in applications like military operations, search-and-rescue (SAR) missions, and disaster relief, where secure and undetectable communication is critical in dynamic or hostile environments.

\section{System Model}
\label{sec: model}

\subsection{General Network Model}

Consider a network of $N$ system nodes ${\cal T}=\{T_1, T_2, \ldots, T_N\}$.  Each node has $M$ transmission modes available. The goal is to efficiently transmit data from source node $S$ to destination node $D$, potentially via multi-hop transmission without detection of that transmission by an attentive and capable adversary named ``Willie.'' We assume Willie's location is known because he operates from a fixed or previously-localized infrastructure (e.g., visible listening tower) which practical reconnaissance can identify. Consistent with the standard covert communication model \cite{sobers2017covert, lee2018covert}, we assume that the channel state information (CSI) is known to both  Willie and the legitimate nodes. Also present is one jammer ($J$) whose presence results in random interference at the nodes.  As described in \cite{sobers2017covert}, it may be a ``friendly'' jammer in the environment focused on facilitating covert communications, or it may simply be a node transmitting for other purposes.  Each transmission on mode $m$ on each network link is subject to multipath fading, additive white Gaussian noise (AWGN), and pathloss $\alpha_m$ that depends on the radio technology of mode $m$.

A path in the multi-hop network will consist of a sequence of links; on link $i$, the link source $S_i$ attempts to transmit information to the link destination $D_i$.  As in \cite{bash2013limits}, standard Gaussian codebooks of length $n$ will be employed on each link under each mode; in other words, messages are transmitted by node $S_i$ on mode $m$ by drawing a sequence from a collection of randomly constructed length-$n$ sequences of independent and identically distributed (i.i.d.) zero-mean unit-variance Gaussian random variables.  Denote the transmission on mode $m$ by $[f_{m,1}, f_{m,2},\ldots, f_{m,n}]$.  

\begin{figure}[htp]
\centering
\includegraphics[scale=0.3]{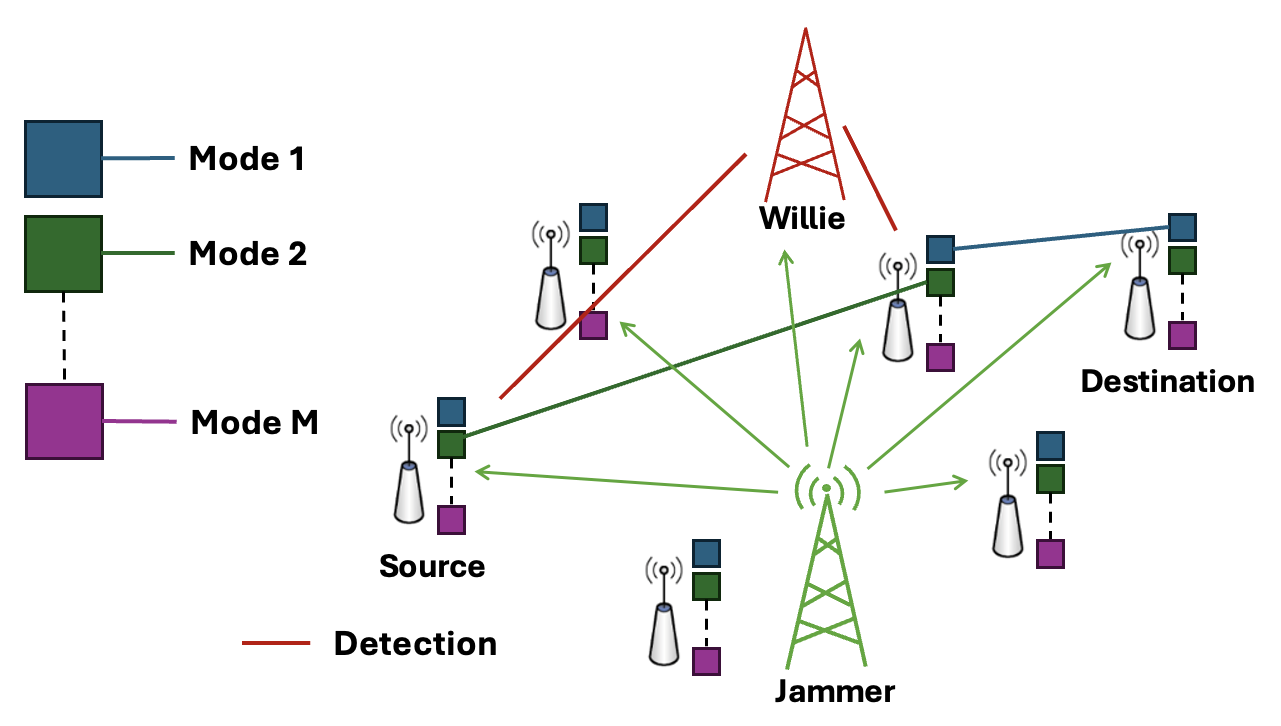}
\caption{Proposed network model: Source to destination communication via friendly relay nodes, each supporting multiple transmission modes (shown in different colors). A friendly jammer injects interference, while an adversary (“Willie”) monitors all transmissions. Red links highlight potential detection paths.}
\label{fig:model}
\end{figure}

Per above, also present in the environment is a ``friendly'' jammer that may decide to transmit random noise to help hide the transmission or whose signal causes a similar effect (e.g., a constant power interferer subject to fading).  The friendly jammer operates in the following manner \cite{sobers2017covert}:  for each code block of length $n$, the jammer draws its power $P_{J,m}, m=1,2,\ldots,M$ independently and uniformly from the interval $[0,\overline{P}_{J,m}]$ and then, for each $m$, $m=1,2,\ldots,M$, transmits random signals $[j_{m,1}, j_{m,2},\ldots, j_{m,n}]$, drawn from a Gaussian distribution with zero-mean and variance $P_{J,m}$. The overall system setup is illustrated in Figure \ref{fig:model}.

The signal from the \(k^{th}\) symbol that (friendly) receiver $D_i$ receives on mode $m$ for $k=1,2,\ldots,n$ is given by:
\begin{equation*}
    \begin{split}
    Z_{m,k}^{(D_i)} = \frac{g_{(S_i,D_i),m} \sqrt{P_{S_i,m}}}{d_{S_i,D_i}^{\alpha_m/2}} f_{m,k} + \frac{g_{(J,D_i),m} \sqrt{P_{J,m}}}{d_{J,D_i}^{\alpha_m/2}} j_{m,k} + N^{(D_i)}_{m,k},
    \end{split}
\end{equation*}
where $g_{(S_i,D_i),m}$ is the fading gain between transmitter $S_i$ and receiver $D_i$ on mode $m$ for link i, $P_{S_i,m}$ is the transmit power of node $S_i$ on mode $m$, $d_{S_i,D_i}$ is the distance between node $S_i$ and $D_i$, and $N^{(D_i)}_{m,i}\sim {\cal N}(0,\sigma_{D_i,m}^2)$ is the AWGN at the receiver for mode $m$ at node $D_i$.  The signal that Willie $W$, receives on mode $m$ is given for $k=1,2,\ldots,n$ by:
\begin{equation*}
\begin{split}
    Z_{m,k}^{(W)} = \frac{g_{(S_i,W),m} \sqrt{P_{S_i,m}}}{d_{S_i,W}^{\alpha_m/2}} f_{m,k} + \frac{g_{(J,W),m} \sqrt{P_{J,m}}}{d_{J,W}^{\alpha_m/2}} j_{m,k} + N^{(W)}_{m,k},
    \end{split}
\end{equation*}
where $g_{(X,W),m}$ and $d_{X,W}$ are defined analogously to above, and $N^{(W)}_{m,i} \sim {\cal N}(0,\sigma_{W,m}^2)$.

\subsection{Adversary Model and Covertness Definition}
\subsubsection{Adversary Model}
The adversary Willie will attempt to detect whether there is communication from a source to the destination in the network. To consider a strong adversary, we assume: (1) the adversary knows the parameters of each link in the network, including the power employed by each communicating node on each mode; (2) the adversary is aware of the jammer's location and its power distribution on each mode (i.e., the values of $\overline{P}_{J,m}$ for $~m=1,2,\ldots,M$); (3) the adversary observes all modes across all links and makes an optimal decision based on those observations, hence ensuring that our approaches satisfy the true end-to-end covertness constraint pertinent to network communications.  Note that friendly transmitting and receiving nodes have access to the codebooks, which are unknown to the adversary, thereby providing the strategic advantage for covert communication \cite{bash2013limits}.  

\subsubsection{Covertness}

Define $H_0$ as the hypothesis that there is no communication between $S$ and $D$, and $H_1$ as the hypothesis that there is such a transmission.  In trying to choose between these hypotheses based on his observations, the probability of error at Willie under equally likely hypotheses is given by \(P_E = P_{FA} + P_{MD}\), where \(P_{FA}\) represents the probability of false alarm, and \(P_{MD}\) represents the probability of missed detection.  If Willie flips a coin, then $P_E = P_{FA} + P_{MD} = 1$, and hence the definition of covertness at level $\epsilon > 0$ is that $P_{FA} + P_{MD} \geq 1 - \epsilon$ \cite{bash2013limits}.
\section{Adversary Analysis}
\label{sec:adversary_analysis}

As in many covert communication analyses, evaluating the performance of the adversary's receiver is critical and challenging.  From the $n$-length observations across modes $m=1,2,\ldots,M$, the adversary Willie seeks to determine whether $S$ transmitted ($H_1$) or not ($H_0$) to $D$. 

Consider first the $i^{th}$ link, with source $S_i$ and destination $D_i$, along the path from $S$ to $D$.  It can be shown that the optimal detector at the adversary is identical to that of trying to detect a transmission from $S_i$ in $M$-block fading, and the efficient analysis of such a receiver is a longstanding unresolved challenge in the covert communications literature \cite{sobers2017covert,spawc2018}.
Instead, we upper bound the performance of the adversary's receiver so as to provide covertness guarantees.  First, define the average power received at the adversary during a given codeword in mode $m$ as:
\begin{equation}
\gamma_m=\frac{g^2_{(S_i,W),m} P_{S_i,m}}{d_{S_i,W}^{\alpha_m}} + \frac{g^2_{(J,W),m} P_{J,m}}{d_{J,W}^{\alpha_m}} + \sigma_{W,m}^2
\end{equation}
Then, the relation of the active hypothesis to the average power received and observations can be expressed as
\begin{equation*}
\begin{split}
H_p \longrightarrow {\underline \gamma}=[\gamma_1, \gamma_2, \ldots, \gamma_M]^T \longrightarrow \{Z_{m,k}^{(W)}, m=1,2,\ldots, M; 
\\
k=1,2,\ldots,n, \quad p=0,1\},
\end{split}
\end{equation*} 
where $X \longrightarrow Y \longrightarrow Z$ means that $X$ and $Z$ are conditionally independent given $Y$.  Hence, a genie-aided receiver for the adversary Willie that provides to him ${\underline \gamma}$ has performance that upper bounds the performance of his actual receiver (i.e., is pessimistic from the point of view of the network).

Given a transmission ($H_1$) or not ($H_0$), the only unknowns in the observations $\underline \gamma$ for the genie-aided Willie are the (random) variables $P_{J,m},m=1,2,\ldots,M$.  Willie uses his understanding of the (uniform) distribution of the jamming to attempt to classify $\underline \gamma$ as $H_0$ or $H_1$.  Consider the hyperbox
\begin{eqnarray}
{\mathcal B}=\left [\frac{g^2_{(S_i,W),1} P_{S_i,1}}{d_{S_i,W}^{\alpha_1}}, \frac{g^2_{(J,W),1} \overline{P}_{J,1}}{d_{J,W}^{\alpha_1}} \right ] 
\nonumber \\
\times  \left [\frac{g^2_{(S_i,W),2} P_{S_i,2}}{d_{S_i,W}^{\alpha_2}}, \frac{g^2_{(J,W),2} \overline{P}_{J,2}}{d_{J,W}^{\alpha_2}} \right] \nonumber \\
\times \cdots \times \left [\frac{g^2_{(S_i,W),M} P_{S_i,M}}{d_{S_i,W}^{\alpha_m}}, \frac{g^2_{(J,W),M} \overline{P}_{J,M}}{d_{J,W}^{\alpha_M}} \right].
\end{eqnarray}
By the law of total probability, 
\begin{eqnarray}
    P_E & = & P_{E|{\mathcal B}} \cdot P(\underline{\gamma} \in {\mathcal B}) + P_{E|\overline{\mathcal B}} \cdot (1 - P(\underline{\gamma} \in {\mathcal B})) 
    \label{law_of_total}
\end{eqnarray}
\no where $P_{E|{\mathcal B}}$ is the adversary's probability of error given $\underline{\gamma} \in {\mathcal B}$, and $P_{E|\overline{\mathcal B}}$ is the adversary's probability of error given $\underline{\gamma} \notin {\mathcal B}$. Since the second term in (\ref{law_of_total}) is positive, 
\begin{eqnarray}
    P_E \geq P_{E|{\mathcal B}} \cdot P({\mathcal B})
\end{eqnarray}
Now, given the (uniform) jamming construction, if $\underline \gamma \in {\mathcal B}$, then $P(H_0 | {\underline \gamma}) = P(H_1 | {\underline \gamma})$, and thus the adversary Willie has performance that is equivalent to flipping a coin when $\gamma \in {\mathcal B}$; in other words, $P_{E|{\mathcal B}}=1$.  This implies that $P_E$ is lower bounded by $P(\underline{\gamma} \in {\mathcal B})$, where:
\begin{eqnarray}
P(\underline{\gamma} \in \mathcal B) & = & \prod_{m=1}^M \frac{\frac{g^2_{(J,W),m} \overline{P}_{J,m}}{d_{J,W}^{\alpha_m}} - \frac{g^2_{(S_i,W),m} P_{S_i,m}}{d_{S_i,W}^{\alpha_m}}}{\frac{g^2_{(J,W),m}\overline{P}_{J,m}}{d_{J,W}^{\alpha_m}}} \\
& = & \prod_{m=1}^M \left (1-\frac{g^2_{(S_i,W),m} P_{S_i,m} d_{J,W}^{\alpha_m}}{g^2_{(J,W),m}\overline{P}_{J,m} d_{S_i,W}^{\alpha_m}}\right ) \\
& \geq & 1 - \sum_{m=1}^M \frac{g^2_{(S_i,W),m} P_{S_i,m} d_{J,W}^{\alpha_m}}{g^2_{(J,W),m}\overline{P}_{J,m} d_{S_i,W}^{\alpha_m}}
\end{eqnarray}
where the last inequality is true if each of the summands is less than one, which will be naturally enforced in the following.  Hence, a covertness constraint of $P(E) \geq 1 - \epsilon_i$ on link $i$ is enforced if:
\begin{eqnarray}
\sum_{m=1}^M \frac{g^2_{(S_i,W),m} P_{S_i,m} d_{J,W}^{\alpha_m}}{g^2_{(J,W),m}\overline{P}_{J,m} d_{S_i,W}^{\alpha_m}} < \epsilon_i.
\label{eq:covert_constraint_per_link}
\end{eqnarray}
and, when the adversary observes links $i=1,2,\ldots,L$, an end-to-end covertness constraint of $P(E) \geq 1 - \epsilon$ is enforced if:
\begin{eqnarray}
\sum_{i=1}^L \sum_{m=1}^M \frac{g^2_{(S_i,W),m} P_{S_i,m} d_{J,W}^{\alpha_m}}{g^2_{(J,W),m}\overline{P}_{J,m} d_{S_i,W}^{\alpha_m}} < \epsilon.
\label{eq:covert_constraint}
\end{eqnarray}
\externaldocument{Adversary_Analysis}
\section{Proposed Algorithms}
\label{sec: Proposed Algorithms}
Here we propose algorithms for jointly optimal routing and link configuration that maximize the end-to-end data rate in a heterogeneous network in the presence of a jammer, while being subject to a covertness constraint against an attentive and powerful adversary who observes all of the links, as described in \ref{sec: model}. 
\subsection{Linearized Covert Optimal Routing in the presence of a Jammer (L-CROP-J)}
\label{subsec: L-CROP-J}
The low covert throughput under strict covertness constraints motivates linearizing the capacity expression for a single link, as done in \cite{gillani2024optimal}.  Note that this linearization will be removed in subsequent sections.

\textbf{Single-link analysis:}
We consider the optimal combination of modes to be used to maximize the capacity of a single link subject to the covertness constraint (\ref{eq:covert_constraint_per_link}).  We assume that the rate from $S_i$ to $D_i$ is chosen so that the communication is reliable, even when the jammer is operating at its highest power. Then, the optimization problem is:
\begin{eqnarray}
\displaystyle \mbox{max}_{\{P_{S_i,1}, P_{S_i,2},\ldots, P_{S_i,M}\}}
\sum_{m=1}^M \log_2 \left (1 + \frac{ \frac{g^2_{(S_i,D_i),m} P_{S_i,m}}{d_{S_i,D_i}^{\alpha_m}}}{\sigma^2_{D_i,m} + \frac{g^2_{(J,D_i),m}\overline{P}_{J,m}} {d_{J,D_i}^{\alpha_m}} } \right )
\label{eq:capacity1}
\end{eqnarray}
subject to:
\begin{eqnarray}
\sum_{m=1}^M \frac{g^2_{(S_i,W),m} P_{S_i,m} d_{J,W}^{\alpha_m}}{g^2_{(J,W),m}\overline{P}_{J,m} d_{S_i,W}^{\alpha_m}} < \epsilon_i
\label{eq:constraint}
\end{eqnarray}
As noted above, we anticipate that at this point, covert capacities will be small, and hence, we linearize the capacity expression to arrive at the optimization problem:
\begin{eqnarray}
\mbox{max}_{\{P_{S_i,1},\ldots, P_{S_i,M}\}}\sum_{m=1}^M \frac{ \frac{g^2_{(S_i,D_i),m} P_{S_i,m}}{d_{S_i,D_i}^{\alpha_m}}}{\sigma^2_{D_i,m} + \frac{g^2_{(J,D_i),m}\overline{P}_{J,m}} {d_{J,D_i}^{\alpha_m}} }
\label{eq:linearized_capacity}
\end{eqnarray}
subject to $(\ref{eq:constraint})$.  We have a linear objective with linear constraints.  Thus, the maximum occurs at a vertex of the constraint space, implying that, in contrast to \cite{gillani2024optimal}, mode selection is optimal for a given link under the assumption of a single jammer.  Letting
\begin{eqnarray}
a_m =  \frac{ \frac{g^2_{(S_i,D_i),m}}{d_{S_i,D_i}^{\alpha_m}}}{\sigma^2_{D_i,m} + \frac{g^2_{(J,D_i),m}\overline{P}_{J,m}} {d_{J,D_i}^{\alpha_m}} } ~~~~~
b_m =  \frac{g^2_{(S_i,W),m} d_{J,W}^{\alpha_m}}{g^2_{(J,W),m}\overline{P}_{J,m} d_{S_i,W}^{\alpha_m}}
\label{eq:constants}
\end{eqnarray}
One should select the mode such that $a_m/b_m$ is maximized.  Lumping multiplicative terms that do not depend on $m$ into a constant $K$ yields that we should select the mode $m$ for which
\begin{eqnarray}
    \frac{a_m}{b_m} = K \frac{g^2_{(J,W),m} g^2_{(S_i,D_i),m} \overline{P}_{J,m}}{(\sigma^2_{D_i,m} d_{J,D_i}^{\alpha_m} + g^2_{(J,D_i),m}\overline{P}_{J,m}) \cdot g^2_{(S_i,W),m} },
    \label{eq:best_mode}
\end{eqnarray}
is the largest.  
As expected, modes are more favorable if: (i) the jammer to adversary and source to destination gains are large; (ii) the jammer power available is large; (iii) there is less noise power at the destination; and (iv) the jammer to receiver and source to adversary gains are smaller.

\textbf{Routing:} Using the single link results, each potential link $i$ uses the mode $m(i)$ that maximizes $a_m/b_m$.
Then, the per-link capacity can be written as:
$C_i(\epsilon_i) = \epsilon_i \beta_i$.
Here, $\beta_i$ depends only on the link parameters and the selected mode:
\begin{eqnarray}
\label{eq:beta_i}
\beta_i = \frac{g^2_{(S_i,D_i),m(i)} \overline{P}_{J,m(i)} g^2_{(J,W),m(i)} d^{\alpha_m}_{S_i,W} d^{\alpha_m}_{J,D_i}}
    {(d^{\alpha_m}_{J,D_i} + \sigma_{D_i,m(i)}^2) g^2_{(S_i,W),m(i)} d^{\alpha_m}_{J,W} d^{\alpha_m}_{S_i,D_i}},
\end{eqnarray}
and
\begin{eqnarray}
\label{eq:power}
     P_{i,m(i)}
     = \epsilon_i\frac{\overline{P}_{J,m(i)}d^{\alpha_m}_{S_i,W} g^2_{(J,W),m(i)}}{d^{\alpha_m}_{J,W}g^2_{(S_i,W),m(i)}}.
\end{eqnarray}
Consider that the adversary observes the collection of signals from all of the links as a message moves through the network.  Per \eqref{eq:covert_constraint}, an end-to-end covertness constraint is maintained for a path with $L$ hops if: $\sum_{i=1}^L \epsilon_i \leq \epsilon$.
Hence, for a given path \(\Pi\), our goal is to maximize the capacity over the allocation of the covertness constraint; that is,
\begin{eqnarray}
    \mbox{max}_{\{\epsilon_1,\epsilon_2,\ldots,\epsilon_L\}} C_{\Pi} = \mbox{max}_{\{\epsilon_1,\epsilon_2,\ldots,\epsilon_L\}} ~~\mbox{min}_{i=1,2,\ldots L} ~ C_i(\epsilon_i)
\end{eqnarray}
such that end-to-end covertness constraint $\sum_{i=1}^L \epsilon_i \leq \epsilon$ is satisfied.  Since the end-to-end capacity is limited by the lowest capacity link, the capacity of the links along the path is equal at the optimal point. Mathematically:
\[C_1(\epsilon_1)=C_2(\epsilon_2)=\cdots= C_L(\epsilon_L)=C_\Pi\]
Substituting into the end-to-end covertness constraint yields:
\begin{eqnarray}
\sum_{i=1}^L \frac{C_{\Pi}}{\beta_i} = C_{\Pi} \sum_{i=1}^L \frac{1}{\beta_i} \leq \epsilon
\end{eqnarray}
and thus
\begin{eqnarray}
\label{eq:optimal_capacity_LCROPJ}
C_{\Pi} = \frac{\epsilon}{\sum_{i=1}^L \frac{1}{\beta_i}}
\end{eqnarray}
is the capacity of the route.  Hence, the best route is that for which $\sum_{i=1}^L \frac{1}{\beta_i}$ is minimized, which is easily found in polynomial time via a shortest path algorithm with weight (cost) of $\frac{1}{\beta_i}$ per link as presented in Algorithm \ref{alg: L-CROP-J}.
\RestyleAlgo{ruled}
\begin{algorithm}[t]
\caption{({\bf L-CROP-J}) Linearized Covert Optimal Routing algorithm with Jammer with a runtime of \(O(N^2)\) for \(N\) nodes}\label{alg: L-CROP-J}
    1. For each pair of nodes in the network, compute \(\frac{a_m}{b_m}\) as given in \eqref{eq:best_mode} for that potential link.

    2. For each potential link in the network, select the optimal mode of operation by selecting the mode with the maximum value of \(\frac{a_m}{b_m}\).
    
    3. Compute \(\beta_i\) for the potential link between each pair of nodes in the network using \eqref{eq:beta_i}.
    
    4. Find the optimal route $\Pi^*$ that minimizes \(\sum_{i=1}^L \frac{1}{\beta_i}\), using Dijkstra's algorithm \cite{Dijkstra1959} while setting the weight of each link as \(\frac{1}{\beta_i}\). 

    5. Compute the capacity $C_{\Pi^*}$ for the optimal path using \eqref{eq:optimal_capacity_LCROPJ}.
    
    6.  For each link $l_i$, $i=1,2,\ldots,L$ in the optimal path $\Pi=(l_1,l_2,\ldots, l_L)$ returned from Dijkstra's algorithm:
    
    \begin{itemize}
        \item {Calculate the link's (optimal) covertness constraint \\ 
        with $\epsilon_i= \frac{C_{\Pi^*}}{\beta_i}$.}
        \item{Use $\beta_i$, $\epsilon_i$, and the link parameters to find the  \\ optimal power 
        \(P_{i,m(i)}\) for link $l_i$ 
        using \eqref{eq:power}.}
    \end{itemize}
\end{algorithm}

\subsection{Mode Selection Covert Optimal Routing algorithm with Jammer (S-CROP-J)}

While L-CROP-J relies on the low data-rate approximation $C=\log_2(1+x)\approx x$, the presence of a jammer can increase achievable rates, motivating retention of the logarithmic capacity expression. 
In many practical scenarios, mode selection is preferable to full power allocation across all modes due to its significantly lower computational complexity or because of reduced interference in a multi-flow environment.   Previous works \cite{kong2024covert, kong2025joint,  kong2024decentralized} often assume mode selection from the onset for these reasons. With this motivation, here we adopt a mode selection strategy in the presence of a jammer, which also allows us to find a polynomial-time jointly optimal routing and link configuration while employing the true (non-linearized) capacity expression for each link.  

Since only the optimal mode $m(i)$ is selected for each link in the presence of the jammer, the key term in the covertness constraint depends only on the link parameters and the selected mode, given by $\beta_i$ from \eqref{eq:beta_i}. The corresponding per-link capacity is expressed as:
\begin{eqnarray}
C_i(\epsilon_i) = \log_2(1+\epsilon_i \beta_i).
\label{eq:C_terms_epsilon_i}
\end{eqnarray}
At the optimal point, all link capacities are equal, i.e.,
\begin{eqnarray}
C_\Pi = C_i(\epsilon_i) = \log_2(1+\epsilon_i \beta_i).
\end{eqnarray}
This implies $\epsilon_i = \frac{2^{C_\Pi}-1}{\beta_i}$,
with the end-to-end constraint $\sum_{i=1}^L \epsilon_i = \epsilon.$ Substituting the above expression for 
$\epsilon_i$ yields the following end-to-end capacity:
\begin{eqnarray}
C_\Pi = \log_2 \left(1+\frac{\epsilon}{\sum_{i=1}^L \frac{1}{\beta_i}}\right).
\label{eq:S-CROP-J_capacity}
\end{eqnarray}
The optimal route under the accurate capacity formulation, therefore, depends on minimizing $\sum_{i=1}^L1/{\beta_i}$, similar to the L-CROP-J case, but the explicit logarithmic dependence leads to a more accurate representation at moderate data rates. Moreover, the optimal distribution of the covertness parameters per-link $\epsilon_i$ differs between the accurate (S-CROP-J) and approximate (L-CROP-J) models. We will see in \ref{sec: Numerical results} that the end-to-end covert throughputs of L-CROP-J and S-CROP-J are very similar at lower capacities, and S-CROP-J performs better at higher throughputs; therefore, we can conclude that the per-link $\epsilon_i$ differs significantly.
\begin{algorithm}[h]
\caption{({\bf S-CROP-J}) Determines the optimal route and per-link covertness using accurate capacity expression with mode selection in presence of a jammer.}
\label{alg:S-CROP-J}

1. Follow steps 1-4 of Algorithm \ref{alg: L-CROP-J} to find optimal route $\Pi^*$.

2. Compute the accurate end-to-end capacity
\[
C_{\Pi^*} = \log_2 \left( 1 + \frac{\epsilon}{\sum_{i=1}^L \frac{1}{\beta_i}} \right).
\]

3. For each link $i \in \Pi^*$, compute the per-link covertness
\[
\epsilon_i = \frac{2^{C_{\Pi^*}} - 1}{\beta_i}.
\]

4. For each link, compute the per-link transmitted power using \eqref{eq:power}.

\end{algorithm}
\subsection{Optimal Link configuration and Route Selection via Water-filling (WF-Optimal)}
For comparison purposes, here we employ the accurate capacity formulation and all possible modes for each given link;  this gives us a benchmark algorithm that combines the water-filling-based link configuration and an exhaustive search to find the optimal route from a source and a destination.
First, we formulate the following optimization problem to maximize the capacity of a single link under the covertness constraint by assuming $\tilde{P}_m=b_m P_m$:
\begin{eqnarray}
\max_{\tilde{P}_m} \sum_{m=1}^M \log_2\left(1+ \frac{a_m}{b_m} \tilde{P}_m\right)
\end{eqnarray}
subject to
\begin{eqnarray}
\sum_{m=1}^M \tilde{P}_m \le \epsilon_i.
\end{eqnarray}

The corresponding Lagrangian is
\begin{eqnarray}
    \mathcal{L}((\tilde{P_1}, \tilde{P_2}, \dots, \tilde{P_m}), \lambda)= \sum_{m=1}^M \log_2 \left(1+\frac{a_m}{b_m}\tilde{P}_m\right)
\end{eqnarray}
Differentiating with respect to $\tilde{P_m}$ yields the classical water-filling solution:
\begin{eqnarray}
\tilde{P}m = \left(\psi - \frac{b_m}{a_m}\right)^+,
\end{eqnarray}
where 
$\psi= \tilde{P}_m+\frac{b_m}{a_m}$ is adjusted so that
\begin{eqnarray}
    \sum_{m=1}^M \left(\psi-\frac{b_m}{a_m}\right)^+=\epsilon_i
    \label{eq:waterfilling-epsilon}
\end{eqnarray}
This allows computation of the per-link capacity:
\begin{eqnarray}
    C_i(\epsilon_i)= \sum_{m=1}^M \log_2 \left(1+ \frac{a_m}{b_m}\tilde{P}_m\right)
    \label{eq:capacity_waterfilling}
\end{eqnarray}
Due to the complex dependence of $C_i(\epsilon_i)$ on $\epsilon_i$, in the accurate expression of the capacity (see (\ref{eq:waterfilling-epsilon}) and (\ref{eq:capacity_waterfilling})), we cannot derive a polynomial-time routing algorithm similar to HetOpt proposed in \cite{gillani2024optimal} or Algorithm \ref{alg: L-CROP-J} (L-CROP-J) or Algorithm \ref{alg:S-CROP-J} (S-CROP-J) in this paper. In the earlier algorithm, the per-link capacity $C_i(\epsilon_i)$ had the simple form given in (\ref{eq:C_terms_epsilon_i}), which led directly to a closed-form structure in (\ref{eq:S-CROP-J_capacity}) and enabled an efficient polynomial-time solution. 
In contrast, $C_i(\epsilon_i)$ results from a water-filling power allocation over modes, and its inverse mapping to find $\epsilon_i$ for a given $C_\Pi$ is only available through an iterative search. Because of this, we cannot obtain a relation for end-to-end capacity that mirrors (\ref{eq:S-CROP-J_capacity}), and therefore, the same polynomial-time approach cannot be extended.

Given this limitation, we resort to an exhaustive search over all feasible routes for this benchmark. For each route, we allocate power across modes using water-filling and iteratively adjust \(C_\Pi\) via binary search until the resulting $\sum_{i=1}^L \epsilon_i=\epsilon$ matches the covertness budget.
The complete procedure is summarized in Algorithm~\ref{alg:WF-Optimal}.

\begin{algorithm}[htp]
\caption{({\bf WF-Optimal}) Route and link configuration for accurate maximum capacity by water filling and using exhaustive search for all possible routes}\label{alg:WF-Optimal}
    1. Enumerate all paths.

    2. For a given path \(L_1,L_2, \dots, L_P\):
    
        \quad    I. Guess \(C_{min}\) and \(C_{max}\).
        
        \quad    II. Compute \(C_\Pi=\frac{C_{max}-C_{min}}{2}\) and update it based on binary search.
            
        \quad    III. Evaluate \(\epsilon_i\) for each link \(l_1,l_2, \dots, l_L \in L_p\) in a given path using Algorithm \ref{alg:WF-Optimal-link}. 
            
        \quad    IV. From \(\epsilon_i\) for each link, find \(\tilde{\epsilon}= \sum_{i=1}^L \epsilon_i\).
            
        \quad    V. If \(\tilde{\epsilon} < \epsilon\), increase \(C_\Pi\) by \(C_{min}=C_\Pi\) and return to  step II. 
        
        \quad \quad        If \(\tilde{\epsilon} > \epsilon\), decrease \(C_\Pi\) by by \(C_{max}=C_\Pi\) and return to  step II.
        
       \quad \quad         If \(\tilde{\epsilon}\approx \epsilon\).
       
       \quad     VI. \(C_\Pi\) is the capacity for path, \(L_p\).
       
    3. Choose the path with the largest capacity \(C_\Pi\)
\end{algorithm}
\begin{algorithm}[htp]
\caption{({\bf Link\_Optimization\_via\_WaterFilling}) For a given link, find the power allocated to each mode by water-filling }\label{alg:WF-Optimal-link}
    Input \(C_\Pi\).
    
    1. Guess a value for \(\psi_{min}\) and \(\psi_{max}\).

    2. Compute \(\psi=\frac{\psi_{min}+\psi_{max}}{2}\) and form \(\tilde{P}_m = \left(\psi-\frac{b_m}{a_m}\right)^+\).

    3. Evaluate \(C_i(\epsilon_i)=\sum_{m=1}^M \log_2 \left(1+ \frac{a_m}{b_m}\tilde{P}_m\right)\).

    4. If \(C_i(\epsilon_i) < C_\Pi\), \(\psi_{min}=\psi\), and return to step 2. 

    \quad If \(C_i(\epsilon_i) > C_\Pi\), \(\psi_{max}=\psi\), and return to step 2. 
    
    \quad If \(C_i(\epsilon_i) \approx  C_\Pi\), go to step 5. 

    5. Evaluate \(\epsilon_i = \sum_{m=1}^M \left(\psi-\frac{b_m}{a_m}\right)^+\).
    
    6. Return \(\epsilon_i\)
\end{algorithm} 
Although this produces the optimal route and mode configuration under the accurate capacity model, the combination of water-filling and route enumeration becomes computationally expensive $O(N!)$, as the number of nodes $N$ increases.

\subsection{Water-filling Covert optimal Routing with Jammer (WF-CROP-J)}
To bridge the performance gap between the low-complexity S-CROP-J algorithm and the WF-Optimal benchmark, we propose a hybrid approach termed WF-CROP-J. The key idea behind this approach is to decouple the route selection and power allocation. Specifically, we leverage the optimal route found when we force mode selection and minimize the metric $\sum_{i=1}^L \frac{1}{\beta_i}$, which can be done in polynomial time.
Once the route $\Pi^*$ is determined, WF-CROP-J applies water-filling along the selected route to find the optimal power allocation across links and modes using Algorithm \ref{alg:WF-Optimal} and \ref{alg:WF-Optimal-link}. 
The computational complexity to determine the route is the same as S-CROP-J of $O(N^2)$ for $N$ friendly nodes. Second, water-filling-based power allocation is applied only on the selected route, which has at most $N-1$ nodes. For each link, the water-filling is implemented using a binary search over water level $\psi$ with a resolution of $e_{min}$ and search range $d_{psi}$, resulting in the complexity $O(\frac{d_{\psi}}{e_{min}})$. In addition, the end-to-end capacity is determined via a binary search as well with a resolution $e_{min}$ and over a range $d_C$, yielding complexity $O(\frac{d_{C}}{e_{min}})$
Thus, the total complexity of the water-filling process along the chosen path scales as \(
O\!\left((N-1)\frac{d_\psi}{e_{\min}}\frac{d_C}{e_{\min}}\right),
\)
which grows linearly with $N$. Since the routing step dominates asymptotically, the overall complexity of WF-CROP-J is
\(
O(N^2),
\)
making it computationally efficient while achieving performance strictly between S-CROP-J and WF-Optimal. The complexity and routing methods of the four algorithms are summarized in Table \ref{tab:summary}.

\begin{table}[t]
\centering
\caption{Summary of routing strategies and computational complexity of the proposed and benchmark algorithms.}
\scriptsize
\scriptsize
\begin{tabularx}{\linewidth}{|c|X|X|c|}
\hline
\textbf{Algorithm} & \textbf{Assumptions} & \textbf{Key Features} & \textbf{Complexity} \\
\hline
\textbf{L-CROP-J} & Low-rate (linearized) capacity model on each link & Mode selection per link is optimal; Dijkstra routing. & $O(N^2)$ \\
\hline
\textbf{S-CROP-J} & Accurate moderate-rate model retaining logarithmic capacity; a single mode employed per link & Dijkstra routing and accurate allocation of $\epsilon$ along the route. & $O(N^2)$ \\
\hline
\textbf{\makecell{WF-Optimal \\ (Benchmark)}} & Optimal mode combination and route selection (for benchmarking) & Exhaustive route search with water-filling over modes. & $O(N!)$ \\
\hline
\textbf{WF-CROP-J} & Route selection via L-CROP-J or S-CROP-J; optimal allocation along the chosen route & Dijkstra routing followed by water-filling along the selected route. & $O(N^2)$ \\
\hline
\end{tabularx}
\label{tab:summary}
\end{table}
\section{Numerical Results}
\label{sec: Numerical results}

In this section, we present the simulation results to evaluate the performance of our proposed methods compared to existing approaches, including the Het-Opt method introduced in \cite{gillani2024optimal} that performs jointly optimal routing and link allocation without jamming.

\paragraph{Simulation Setup}  We consider networks with $N=8$ to $35$ nodes, one adversary (Willie), and one friendly jammer uniformly placed in a $100\times100$ region. The source and destination are fixed at the lower left and upper right corners, respectively, while Willie, Jammer, and the other friendly nodes are placed randomly. The block length is $n=250$ with a covertness constraint $\epsilon=0.01$.  Three modes are defined:  Mode 1 (AWGN channel), with path-loss exponent $\alpha_1=2$ and channel noise statistics \(\sigma_{w,1}^2 \sim \mathcal{U}(6,8)\) and \(\sigma_{D_i,1}^2 \sim \mathcal{U}(8,10)\); Modes 2 and 3 (fading channels), for which the path-loss exponents are $\alpha_2=4$ and $\alpha_3=3$, and the channel noise statistics are: \(\sigma_{w,2}^2 \sim \mathcal{U}(1,3)\), \(\sigma_{w,3}^2 \sim \mathcal{U}(2,4)\), \(\sigma_{D_i,2}^2 \sim \mathcal{U}(1,2)\) and \(\sigma_{D_i,3}^2 \sim \mathcal{U}(5,7)\), where $\sim \mathcal{U}(a,b)$ means the random variable is uniformly distributed between $a$ and $b$.  This selection provides adversary Willie with an advantage and considers variations in receiver quality at friendly nodes.   For modes 2 and 3, a standard normalized Rayleigh channel fading model is employed; that is, \(g_{(S_i, D_i),m} \sim \mathcal{CN}(0, 1)\) and \(g_{(S_i,W),m} \sim \mathcal{CN}(0, 1)\).  Each plot is an average of \(10^3\) randomly generated networks.  To preserve the randomness of the network topology when the distance between Willie and another node (e.g., the source or the jammer) is fixed, as in some of the simulations below, Willie and the jammer are repeatedly placed at random locations, and only realizations whose pairwise distance falls within the desired range are retained for simulation.

\paragraph{Single-Link Performance} 

\begin{figure}[h]
 \begin{subfigure}{0.232\textwidth}
    \includegraphics[width=\linewidth]{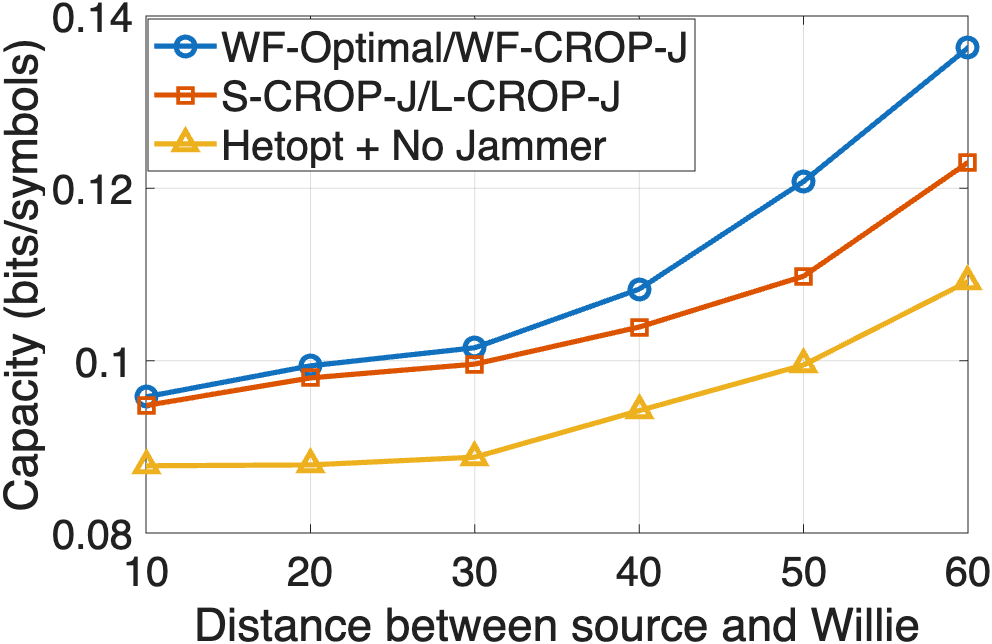}
    \caption{\centering } \label{fig:Single_link_vary}
  \end{subfigure}
  \hspace*{\fill}   
  \begin{subfigure}{0.235\textwidth}
    \includegraphics[width=\linewidth]{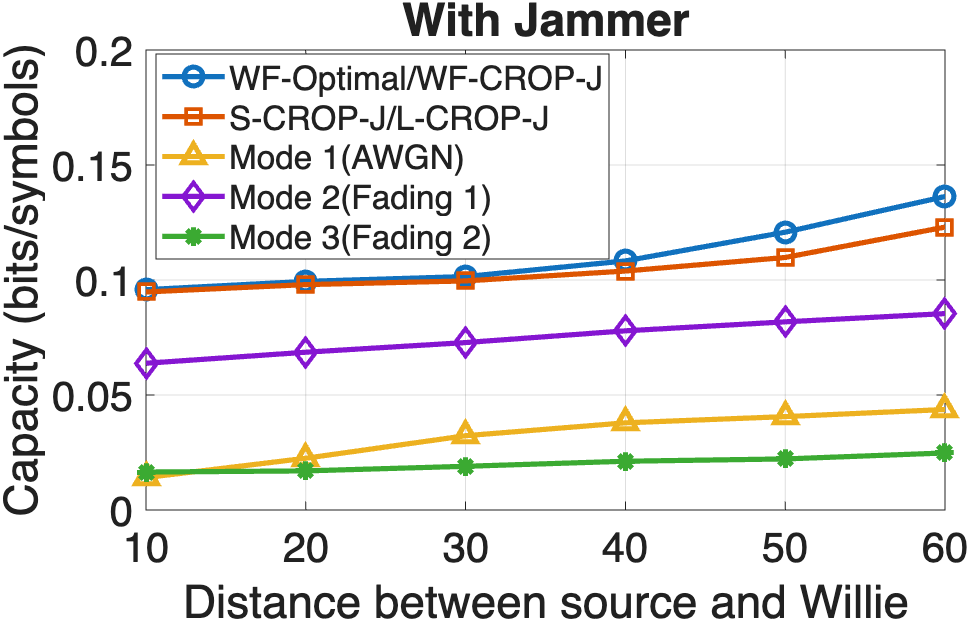}
    \caption{\centering } \label{fig:Single-link_with_jammer}
  \end{subfigure}%
\caption{Single-link covert capacity versus source-Willie distance for the proposed methods (L-CROP-J, S-CROP-J, WF-CROP-J) and the benchmark (WF-optimal). (a) illustrates the performance gain through the use of a friendly jammer, and (b) highlights the improvement obtained by employing multiple modes with the Jammer. The meaningful comparison of the algorithms is considered below in the network scenario. } 
\label{fig:Single-link}
\end{figure}

Figure \ref{fig:Single-link} shows the performance of our proposed methods and compares them with the existing approaches for a single link from a source to the destination in the presence of a Jammer and an adversary Willie. Figure \ref{fig:Single_link_vary} compares the performance of our proposed optimization methods with Het-Opt \cite{gillani2024optimal}.  “HetOpt + No Jammer” represents the performance of [6] in the absence of a jammer. The results indicate that using a jammer can increase the covert capacity significantly. As the distance between the source and Willie increases, the capacity improves because the signal received by Willie weakens due to path loss and other factors, while the jammer is placed randomly. Moreover, as expected, the proposed benchmark water-filling method outperforms all other methods, demonstrating that utilizing a jammer and multiple modes can enhance covert capacity when each mode is used optimally as described in Section \ref{sec: Proposed Algorithms}. Because only a single route exists in the single-link scenario, L-CROP-J and S-CROP-J achieve nearly identical performance, while WF-Optimal and WF-CROP-J coincide. Figure \ref{fig:Single-link_with_jammer} shows the capacity for different methods versus the distance between the source and Willie in the presence of a randomly placed friendly jammer. This result highlights the effectiveness of our proposed method, where multiple modalities are utilized, and the power is allocated optimally across modes, compared to a single mode, as their efficiency can vary depending on the channel characteristics, even in the presence of a jammer.

\paragraph{Joint Routing and Link Optimization}

\begin{figure}[htp]
\centering
\includegraphics[scale=0.35]{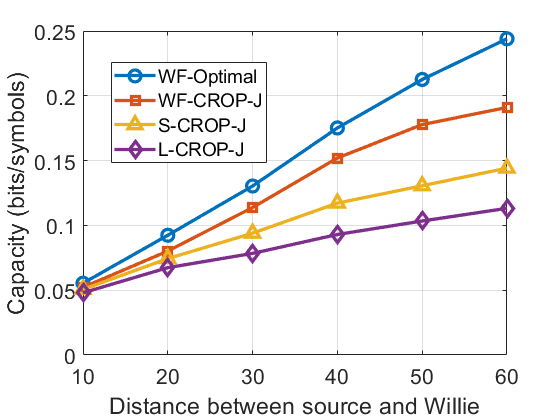}
\caption{Comparison of the end-to-end covert capacity achieved by the proposed routing algorithms (L-CROP-J, S-CROP-J, and WF-CROP-J) and the benchmark exhaustive search method (WF-Optimal) in an eight-node network. The network size is limited to $N=8$ nodes due to the exponential complexity of WF-Optimal. The goal here is to illustrate the performance gap between the low-complexity polynomial-time algorithms and the optimal benchmark in a network setting; a broader scalability comparison is provided in subsequent figures.}
\label{fig:all_4_algos}
\end{figure}

Figure \ref{fig:all_4_algos} shows the end-to-end covert capacity achieved by all four algorithms as a function of the distance between the source and Willie for a network with $N=8$ nodes. When Willie is located close to the source, the achievable capacity is very low for all algorithms, and the performance differences between them are minimal. As the distance between the source and Willie increases, the achievable capacity improves, and the differences between the algorithms become more pronounced. In particular, S-CROP-J achieves slightly higher capacity than L-CROP-J, indicating that the linearized capacity approximation used in L-CROP-J is accurate in the low data-rate regime typical of covert communication, while S-CROP-J provides a preferable alternative when the data rate becomes moderate. As expected, the benchmark exhaustive search method WF-Optimal achieves the highest capacity, while WF-CROP-J consistently outperforms both S-CROP-J and L-CROP-J. Note that WF-Optimal requires an exhaustive search over all possible routes, resulting in exponential computational complexity. In contrast, L-CROP-J, S-CROP-J, and WF-CROP-J are polynomial-time algorithms. Therefore, the network size in this comparison is limited to $N=8$ nodes to make the exhaustive search computationally feasible.

\begin{figure}[htp]
     \begin{subfigure}{0.22\textwidth}
    \includegraphics[width=\linewidth]{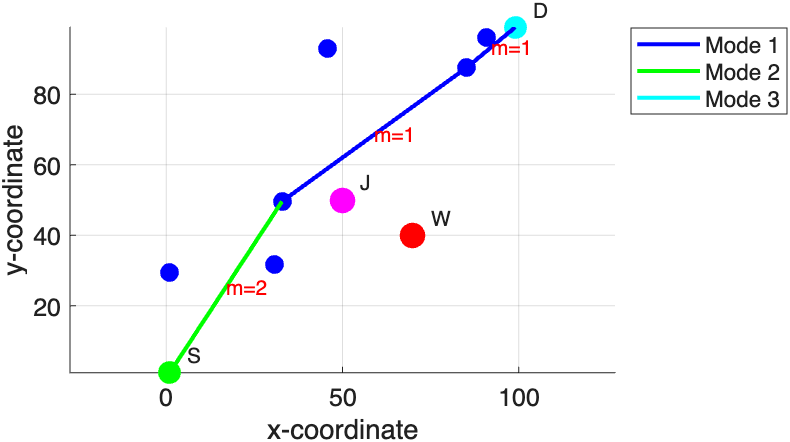}
    \caption{S-CROP-J} \label{fig:S_CROP_J}
  \end{subfigure}
  \hspace*{\fill}  
     \begin{subfigure}{0.22\textwidth}
    \includegraphics[width=\linewidth]{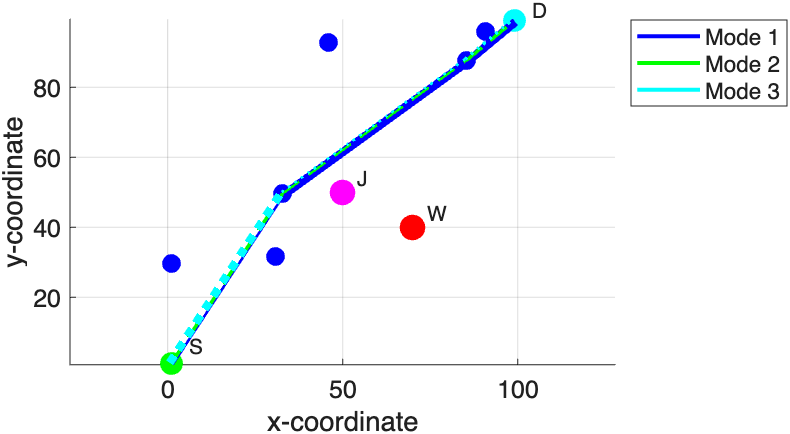}
    \caption{WF-CROP-J} \label{fig:WF_CROP_J}
  \end{subfigure}
    \vspace{0.3cm}
    \centering
     \begin{subfigure}{0.22\textwidth}
    \includegraphics[width=\linewidth]{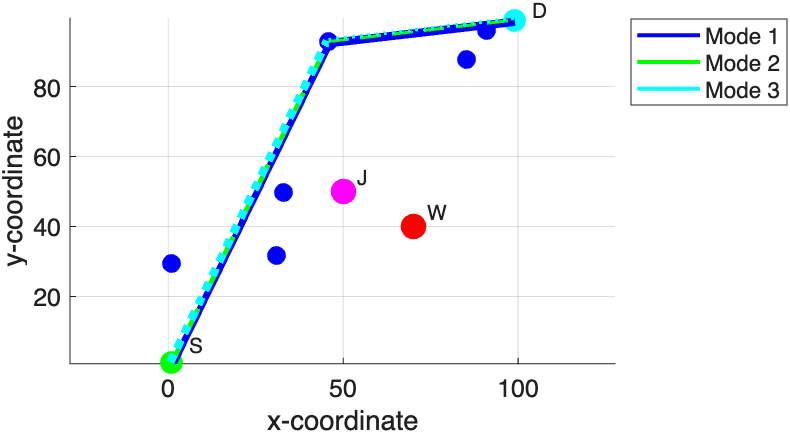}
    \caption{WF-optimal} \label{fig:WF_optimal}
  \end{subfigure}
    \caption{Illustration of the routes selected by S-CROP-J, WF-CROP-J, and the optimal benchmark (WF-Optimal) for a sample network with eight friendly nodes. The goal here is to provide insight into how different routing and power allocation strategies impact path selection and end-to-end covert capacity. The corresponding capacities are $C_\Pi=0.1143$ bits/symbol, $C_\Pi=0.1678$ bits/symbol, and $C_\Pi=0.2126$ bits/symbol, respectively.}
    \label{fig:toy_example}
\end{figure}

\begin{figure}[h]
  \begin{subfigure}{0.235\textwidth}
    \includegraphics[width=\linewidth]{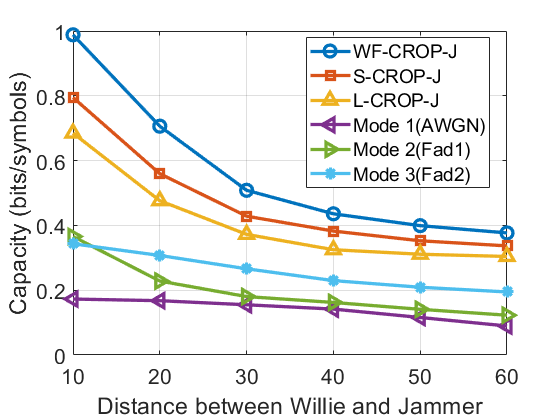}
    \caption{\centering } \label{fig:Routing_distance_between_jammer_and_Willie}
  \end{subfigure}%
  \hspace*{\fill}   
  \begin{subfigure}{0.235\textwidth}
    \includegraphics[width=\linewidth]{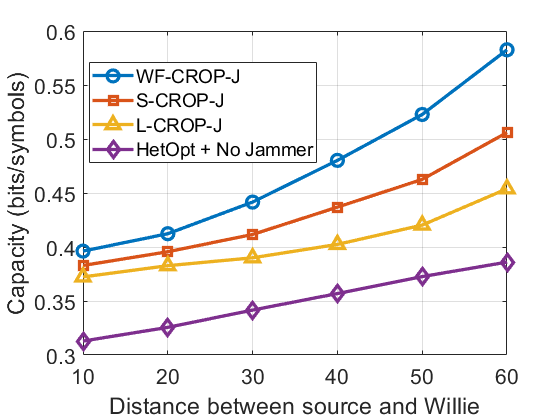}
    \caption{\centering } \label{fig:distance_bw_source_and_Willie}
  \end{subfigure}%
\caption{End-to-end covert capacity achieved by different routing strategies in a network of $N=30$. The goal here is to demonstrate the impact of friendly jamming and multi-mode optimization on covert performance, and to compare the proposed algorithms (L-CROP-J, S-CROP-J, WF-CROP-J) with single-mode operation and Het-Opt without a jammer.}\label{fig:3_algos}
\end{figure}

The utilization of multiple modes is further illustrated in Fig.~\ref{fig:toy_example}. S-CROP-J selects the optimal mode for each link and determines the route that maximizes the covert capacity, resulting in $C_\Pi = 0.1143$ bits/symbol. In contrast, WF-CROP-J uses the same route but performs water filling across all available modes of each link, leading to a higher capacity of $C_\Pi = 0.1678$ bits/symbol. The benchmark algorithm WF-Optimal performs an exhaustive search over all possible routes while optimally allocating power across modes, yielding the highest capacity of $C_\Pi = 0.2126$ bits/symbol.
Fig.~\ref{fig:3_algos} shows the end-to-end covert capacity achieved by the proposed routing algorithms in a network with $N=30$ nodes.  Fig.~\ref{fig:Routing_distance_between_jammer_and_Willie} illustrates the impact of the distance between the jammer and Willie. As the jammer moves farther from Willie, the jamming signal becomes less effective at masking the transmission, resulting in a gradual decrease in covert capacity for all algorithms. In this scenario, WF-CROP-J consistently achieves the highest capacity by optimally allocating power across multiple modes, followed by S-CROP-J and L-CROP-J. The curves corresponding to single-mode operation further illustrate that restricting communication to a single mode significantly limits the achievable covert rate. Fig.~\ref{fig:distance_bw_source_and_Willie} shows the effect of increasing the distance between the source and Willie. As Willie moves farther away from the source, the adversary's detection capability decreases, allowing higher covert transmission rates. Consequently, the covert capacity increases for all routing strategies. The proposed algorithms that exploit friendly jamming and multi-mode optimization outperform the baseline Het-Opt approach without a jammer, with WF-CROP-J again achieving the best performance due to its optimal power allocation across modes.

\begin{figure}[htp]
\centering
\includegraphics[scale=0.35]{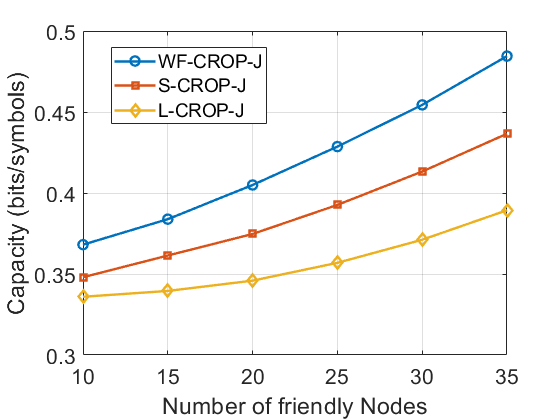}
\caption{Comparison of the end-to-end covert capacity achieved by the proposed routing algorithms (L-CROP-J, S-CROP-J, and WF-CROP-J) for $N=30$. The WF-CROP-J scheme consistently achieves the highest capacity. The capacity increases with the number of friendly nodes due to enhanced routing diversity, which enables the selection of more favorable multi-hop paths and improved covertness.}
\label{fig:cap_vs_N}
\end{figure}

Fig.~\ref{fig:cap_vs_N} illustrates the impact of increasing the number of friendly nodes on the end-to-end capacity achieved by the proposed routing algorithm. As $N$ increases, the network becomes denser, providing a larger set of candidate relay nodes and hence, greater routing diversity. This allows the algorithms to select multi-hop paths that better balance covertness and reliability. WF-CROP-J, consistently achieves better performance than the algorithms L-CROP-J and S-CROP-J.

\begin{figure}[h]
    \centering
     \includegraphics[scale=0.35]{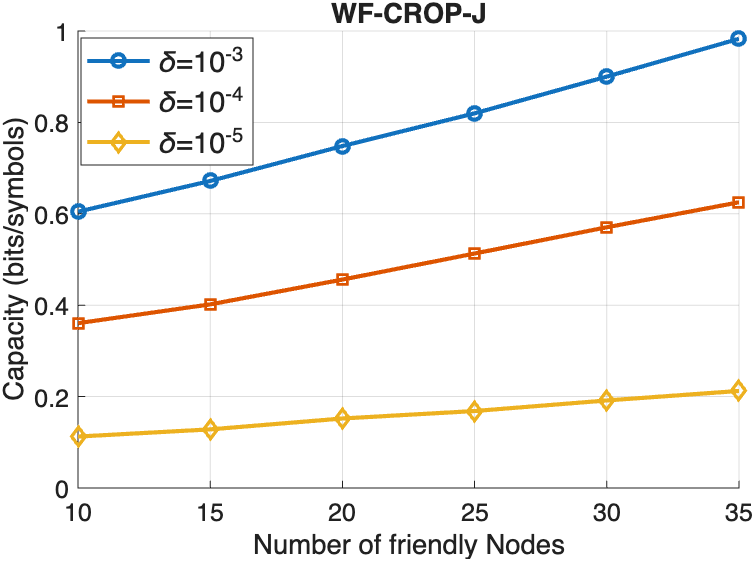}
\caption{Achievable capacity versus the number of friendly nodes for different values of $\delta$. As the covertness constraint becomes more stringent (smaller $\delta$), the achievable capacity decreases, while increasing the number of friendly nodes improves performance across all $\delta$.}\label{fig:vary_delta}
\end{figure}

Fig.~\ref{fig:vary_delta} illustrates this relationship by plotting the achievable capacity versus the number of friendly nodes for different values of $\delta$. For all values of $\delta$, the capacity increases with $N$, reaffirming the benefits of routing diversity. However, the gap between curves corresponding to different $\delta$ values remains substantial, indicating that the covertness constraint plays a dominant role in limiting system performance. Even with a large number of nodes, a stringent covertness requirement (small $\delta$) imposes a hard ceiling on the achievable rate. These results emphasize the fundamental trade-off between covertness and throughput in covert communication systems

\begin{figure}[htp]
\centering
\includegraphics[scale=0.35]{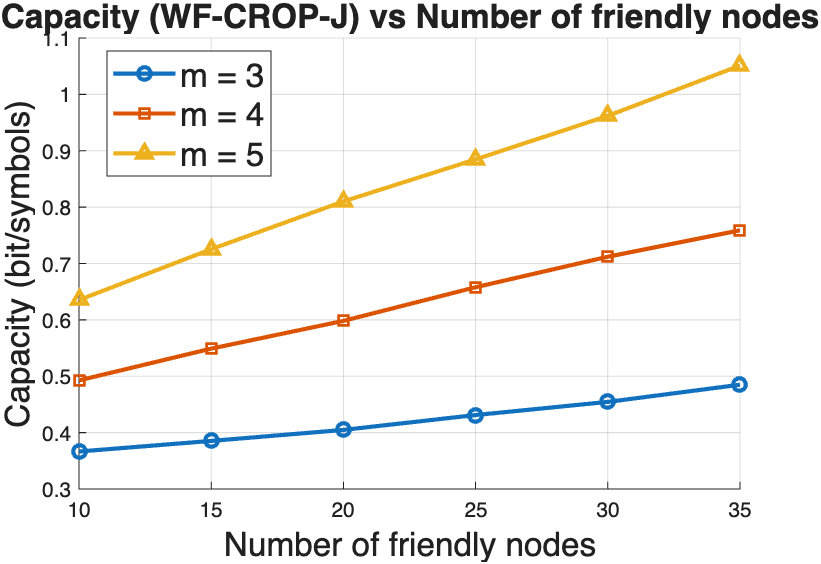}
\caption{Comparison of the end-to-end covert capacity achieved by WF-CROP-J versus the number of friendly nodes for different numbers of modes (m = 3, 4, 5). Increasing the number of modes significantly enhances capacity, and the capacity improves with an increasing number of friendly nodes.}
\label{fig:vary_m}
\end{figure}

Fig.~\ref{fig:vary_m} demonstrates the impact of the number of available transmission modes on the performance of WF-CROP-J. Increasing the number of modes provides additional degrees of freedom for power allocation, enabling the system to better adapt to varying channel conditions across links. As the number of modes increases from $m=3$ to $m=5$, the achievable covert capacity improves significantly across all network sizes. This gain arises because WF-CROP-J performs water-filling over modes, allocating more power to favorable channels while avoiding inefficient ones. With more modes available, the likelihood of encountering high-quality transmission opportunities increases, leading to improved overall performance.

\section{Conclusion}
\label{sec:conclusion}
We considered jointly optimal routing and link configuration in heterogeneous networks (HetNets) in the presence of a jammer. First, we established that per-link mode selection is optimal for low data rates, thereby simplifying the joint routing and link configuration problem.  Leveraging this insight, we propose efficient polynomial-time algorithms that perform route and link optimization. To benchmark performance, we also introduced an exhaustive-search-based approach with optimal power allocation, which provides an upper bound but incurs high computational complexity. Another algorithm with performance between the mode-selection-based approach and the benchmark was introduced, while having lower computational complexity. Numerical results demonstrate the potential for enhancing covert communication in HetNets by introducing a random jammer. An important direction for future work is the extension of the proposed framework to multi-flow scenarios and highly dynamic networks. We also plan to study the practical costs associated with managing multiple active interfaces.

\vspace*{0.2in}

\bibliographystyle{IEEEtran.bst}
\bibliography{ref}

@article{kong2024decentralized,
  title={Decentralized covert routing in heterogeneous networks using reinforcement learning},
  author={Kong, Justin and Moore, Terrence J and Dagefu, Fikadu T},
  journal={IEEE Communications Letters},
  volume={28},
  number={11},
  pages={2683--2687},
  year={2024},
  publisher={IEEE}
}

@ARTICLE{9361424,
  author={Huang, Ke-Wen and Deng, Hao and Wang, Hui-Ming},
  journal={IEEE Transactions on Wireless Communications}, 
  title={Jamming Aided Covert Communication With Multiple Receivers}, 
  year={2021},
  volume={20},
  number={7},
  pages={4480-4494},
  doi={10.1109/TWC.2021.3059306}}

@INPROCEEDINGS{9674386,
  author={Jiang, Junhao and Yang, Weiwei and Ma, Ruiqian},
  booktitle={2021 7th International Conference on Computer and Communications (ICCC)}, 
  title={Joint Relay and Jammer Selection for Covert Communication}, 
  year={2021},
  volume={},
  number={},
  pages={131-135},
  doi={10.1109/ICCC54389.2021.9674386}}

@ARTICLE{10236006,
  author={Kong, Justin and Dagefu, Fikadu T. and Choi, Jihun and Aggarwal, Rahul and Spasojevic, Predrag},
  journal={IEEE Transactions on Vehicular Technology}, 
  title={Covert Communication in Intelligent Reflecting Surface Assisted Networks With a Friendly Jammer}, 
  year={2024},
  volume={73},
  number={1},
  pages={1467-1472},
  doi={10.1109/TVT.2023.3310917}}

@ARTICLE{9120366,
  author={Xiong, Wenhui and Yao, Yinfeng and Fu, Xiaoyu and Li, Shaoqian},
  journal={IEEE Wireless Communications Letters}, 
  title={Covert Communication With Cognitive Jammer}, 
  year={2020},
  volume={9},
  number={10},
  pages={1753-1757},
  doi={10.1109/LWC.2020.3003472}}

@ARTICLE{9360663,
  author={Yang, Bin and Taleb, Tarik and Fan, Yuanyuan and Shen, Shikai},
  journal={IEEE Network}, 
  title={Mode Selection and Cooperative Jamming for Covert Communication in {D2D} Underlaid {UAV} Networks}, 
  year={2021},
  volume={35},
  number={2},
  pages={104-111},
  doi={10.1109/MNET.011.2000100}}

@ARTICLE{9442311,
  author={Zheng, Tong-Xing and Yang, Ziteng and Wang, Chao and Li, Zan and Yuan, Jinhong and Guan, Xiaohong},
  journal={IEEE Transactions on Wireless Communications}, 
  title={Wireless Covert Communications Aided by Distributed Cooperative Jamming Over Slow Fading Channels}, 
  year={2021},
  volume={20},
  number={11},
  pages={7026-7039},
  doi={10.1109/TWC.2021.3080382}}

@ARTICLE{10143419,
  author={Feng, Shaohan and Lu, Xiao and Sun, Sumei and Niyato, Dusit and Hossain, Ekram},
  journal={IEEE Transactions on Wireless Communications}, 
  title={Securing Large-Scale {D2D} Networks Using Covert Communication and Friendly Jamming}, 
  year={2024},
  volume={23},
  number={1},
  pages={592-606},
  doi={10.1109/TWC.2023.3280464}}

@ARTICLE{11214184,
  author={Zhao, Xiang and Lu, Wencong and Yi, Changyan and Wang, Junyi and Peng, Jie},
  journal={IEEE Transactions on Communications}, 
  title={Multi-{UAV} Covert Communication with Informed Jammers: Design, Analysis and Optimization}, 
  year={2025},
  volume={},
  number={},
  pages={1-1},
  doi={10.1109/TCOMM.2025.3624167}}

@ARTICLE{10534096,
  author={Chen, Xun and Gao, Fujun and Qiu, Min and Zhang, Jia and Shu, Feng and Yan, Shihao},
  journal={IEEE Transactions on Information Forensics and Security}, 
  title={Achieving Covert Communication With a Probabilistic Jamming Strategy}, 
  year={2024},
  volume={19},
  number={},
  pages={5561-5574},
  doi={10.1109/TIFS.2024.3402346}}

@ARTICLE{10533214,
  author={Hu, Jinsong and Li, Hongwei and Chen, Youjia and Shu, Feng and Wang, Jiangzhou},
  journal={IEEE Transactions on Wireless Communications}, 
  title={Covert Communication in Cognitive Radio Networks With Poisson Distributed Jammers}, 
  year={2024},
  volume={23},
  number={10},
  pages={13095-13109},
  doi={10.1109/TWC.2024.3398651}}

@ARTICLE{9536435,
  author={Chen, Xinying and Zhang, Ning and Tang, Jie and Liu, Mingqian and Zhao, Nan and Niyato, Dusit},
  journal={IEEE Transactions on Vehicular Technology}, 
  title={{UAV}-Aided Covert Communication With a Multi-Antenna Jammer}, 
  year={2021},
  volume={70},
  number={11},
  pages={11619-11631},
  doi={10.1109/TVT.2021.3112121}}

@INPROCEEDINGS{9834682,
  author={ZivariFard, Hassan and Bloch, Matthieu R. and Nosratinia, Aria},
  booktitle={2022 IEEE International Symposium on Information Theory (ISIT)}, 
  title={Covert Communication in the Presence of an Uninformed, Informed, and Coordinated Jammer}, 
  year={2022},
  volume={},
  number={},
  pages={306-311},
  doi={10.1109/ISIT50566.2022.9834682}}

@ARTICLE{9121687,
  author={Liang, Wei and Shi, Jia and Tie, Zhuangzhuang and Yang, Fucheng},
  journal={IEEE Access}, 
  title={Performance Analysis for {UAV}-Jammer Aided Covert Communication}, 
  year={2020},
  volume={8},
  number={},
  pages={111394-111400},
  doi={10.1109/ACCESS.2020.3001069}}

@ARTICLE{9849051,
  author={Du, Hongyang and Niyato, Dusit and Xie, Yuan-Ai and Cheng, Yanyu and Kang, Jiawen and Kim, Dong In},
  journal={IEEE Journal on Selected Areas in Communications}, 
  title={Performance Analysis and Optimization for Jammer-Aided Multiantenna UAV Covert Communication}, 
  year={2022},
  volume={40},
  number={10},
  pages={2962-2979},
  doi={10.1109/JSAC.2022.3196131}}

@article{sobers2017covert,
  title={Covert communication in the presence of an uninformed jammer},
  author={Sobers, Tamara V and Bash, Boulat A and Guha, Saikat and Towsley, Don and Goeckel, Dennis},
  journal={IEEE TWC},
  volume={16},
  number={9},
  pages={6193--6206},
  year={2017},
  publisher={IEEE}
}

@article{lee2018covert,
  title={Covert communication with channel-state information at the transmitter},
  author={Lee, Si-Hyeon and Wang, Ligong and Khisti, Ashish and Wornell, Gregory W},
  journal={IEEE Transactions on Information Forensics and Security},
  volume={13},
  number={9},
  pages={2310--2319},
  year={2018},
  publisher={IEEE}
}

@ARTICLE{10256033,
  author={He, Rongrong and Chen, Jin and Li, Guoxin and Wang, Haichao and Xu, Yuhua and Yang, Weiwei and Jiao, Yutao and He, Wenhui},
  journal={IEEE Transactions on Vehicular Technology}, 
  title={Channel-Aware Jammer Selection and Power Control in Covert Communication}, 
  year={2024},
  volume={73},
  number={2},
  pages={2266-2279},
  doi={10.1109/TVT.2023.3317638}}

@INPROCEEDINGS{spawc2018,

  author={Goeckel, Dennis and Sheikholeslami, Azadeh and Sobers, Tamara and Bash, Boulat A. and Towsley, Oon and Guha, Saikat},

  booktitle={2018 IEEE 19th International Workshop on Signal Processing Advances in Wireless Communications (SPAWC)}, 

  title={Covert Communications in a Dynamic Interference Environment}, 

  year={2018},

  volume={},

  number={},

  pages={1-5},

  doi={10.1109/SPAWC.2018.8445896}}

@article{gillani2024optimal,
  title={Optimal routing and link configuration for covert heterogeneous wireless networks},
  author={Gillani, Amna and Lorenzo, Beatriz and Ghaderi, Majid and Dagefu, Fikadu and Goeckel, Dennis},
  journal={arXiv preprint arXiv:2412.07059},
  year={2024}
}

@inproceedings{xia2024uav,
  title={UAV-Enabled Covert Cross-Technology Communication in Heterogeneous IoT Networks},
  author={Xia, Xiaohao and Esmat, Haitham H and Lorenzo, Beatriz and Goeckel, Dennis},
  booktitle={2024 IEEE 100th Vehicular Technology Conference (VTC2024-Fall)},
  pages={1--7},
  year={2024},
  organization={IEEE}
}

@article{feng2020smart,
  title="{Smart mode selection using online RL for VR broadband broadcasting in {D2D} assisted  {5G} HetNets}",
  author={Feng, Lei and Yang, Zhixiang and Yang, Yang and Que, Xiaoyu and Zhang, Kai},
  journal={IEEE Transactions on Broadcasting},
  volume={66},
  number={2},
  pages={600--611},
  year={2020},
  publisher={IEEE}
}

@article{algedir2020energy,
  title="Energy efficiency optimization and dynamic mode selection algorithms for {D2D communication under HetNet}",
  author={Algedir, Amal Ali and Refai, Hazem H},
  journal={IEEE Access},
  volume={8},
  pages={95251--95265},
  year={2020},
  publisher={IEEE}
}

@article{xu2021survey,
  title="A survey on resource allocation for {5G HetNets}",
  author={Xu, Yongjun and Gui, Guan and Gacanin, Haris and Adachi, Fumiyuki},
  journal={IEEE Communications Surveys \& Tutorials},
  volume={23},
  number={2},
  pages={668--695},
  year={2021},
  publisher={IEEE}
}

@article{bash2013limits,
  title="Limits of reliable communication with low probability of detection on {AWGN} channels",
  author={Bash, Boulat A and Goeckel, Dennis and Towsley, Don},
  journal={IEEE journal on selected areas in communications},
  volume={31},
  number={9},
  pages={1921--1930},
  year={2013},
  publisher={IEEE}
}

@inproceedings{lee2014achieving,
  title="Achieving positive rate with undetectable communication over {AWGN and Rayleigh channels}",
  author={Lee, Seonwoo and Baxley, Robert J},
  booktitle={2014 IEEE International Conference on Communications (ICC)},
  pages={780--785},
  year={2014},
  organization={IEEE}
}

@article{lee2015achieving,
  title={Achieving undetectable communication},
  author={Lee, Seonwoo and Baxley, Robert J and Weitnauer, Mary Ann and Walkenhorst, Brett},
  journal={IEEE Journal of Selected Topics in Signal Processing},
  volume={9},
  number={7},
  pages={1195--1205},
  year={2015},
  publisher={IEEE}
}

@ARTICLE{covert_routing_2018,

  author={Sheikholeslami, Azadeh and Ghaderi, Majid and Towsley, Don and Bash, Boulat A. and Guha, Saikat and Goeckel, Dennis},

  journal={IEEE Transactions on Wireless Communications}, 

  title={Multi-Hop Routing in Covert Wireless Networks}, 

  year={2018},

  volume={17},

  number={6},

  pages={3656-3669},

  doi={10.1109/TWC.2018.2812881}}

@article{kong2025joint,
  title={Joint routing and resource allocation in covert multi-flow heterogeneous networks},
  author={Kong, Justin H and Moore, Terrence J and Dagefu, Fikadu T},
  journal={IEEE Communications Letters},
  year={2025},
  publisher={IEEE}
}

@article{kong2026simultaneous,
  title={Simultaneous Multi-Modal Covert Communications: Analysis and Optimization},
  author={Kong, Justin H and Moore, Terrence J and Dagefu, Fikadu T},
  journal={arXiv preprint arXiv:2603.12172},
  year={2026}
}

@inproceedings{roknilamouki2025safe,
  title={Safe and Reliable Deep Reinforcement Learning for Covert Routing},
  author={Roknilamouki, Amirhossein and Dagefu, Fikadu T and Ekici, Eylem and Kim, Brian and Kong, Justin and Moore, Terrence and Sun, Yin and Shroff, Ness B},
  booktitle={2025 IEEE 22nd International Conference on Mobile Ad-Hoc and Smart Systems (MASS)},
  pages={37--44},
  year={2025},
  organization={IEEE}
}

@article{kong2024covert,
  title={Covert routing in heterogeneous networks},
  author={Kong, Justin and Dagefu, Fikadu T and Moore, Terrence J},
  journal={IEEE Transactions on Information Forensics and Security},
  volume={19},
  pages={7047--7059},
  year={2024},
  publisher={IEEE}
}

@INPROCEEDINGS{10773635,
  author={Kim, Brian and Kong, Justin and Moore, Terrence J. and Dagefu, Fikadu T.},
  booktitle={MILCOM 2024 - 2024 IEEE Military Communications Conference (MILCOM)}, 
  title={Reinforcement Learning Based Covert Routing with Node Failure Resiliency for Heterogeneous Networks}, 
  year={2024},
  volume={},
  number={},
  pages={709-714},
  doi={10.1109/MILCOM61039.2024.10773635}}

@article{Dijkstra1959,
  author    = {Edsger W. Dijkstra},
  title     = {A Note on Two Problems in Connexion with Graphs},
  journal   = {Numerische Mathematik},
  year      = {1959},
  volume    = {1},
  number    = {1},
  pages     = {269--271},
  doi       = {10.1007/BF01386390}
}

\end{document}